\documentclass[3p,times]{elsarticle}

\usepackage{ecrc}

\volume{00}

\firstpage{1}

\journalname{}

\runauth{Jun-Sik Sin et al}

\jid{}

\jnltitlelogo{}

\usepackage{amssymb}

\usepackage[figuresright]{rotating}

\large
\begin{document}

\begin{frontmatter}



\dochead{}

\title{Ion partitioning effect on the electrostatic interaction between two charged soft surfaces}
\author{\large Jun-Sik Sin}
\large
\ead{js.sin@ryongnamsan.edu.kp}
\address{\large Natural Science Center, \textbf{Kim Il Sung} University, Taesong District, Pyongyang, \\Democratic People's Republic of Korea}

\linespread{1.6}
\begin{abstract}
\large

We theoretically investigate electrostatic properties between two charged surfaces with a grafted polyelectrolyte layer in an aqueous electrolyte solution by using the Poisson-Boltzmann approach accounting for ion partitioning. In order to consider the ion partitioning effect, we focus on changes of electrostatic properties due to the difference in dielectric permittivity of the polyelectrolyte layer and the aqueous electrolyte solution. We find that ion partitioning enhances electrostatic potential in the region between two charged soft surfaces and hence increases the electrostatic interaction between two charged soft surfaces. Ion partitioning effect on osmotic pressure is  enhanced not only by an increase in the thickness of polyelectrolyte layer and Debye length but also by a decrease in ion radius.
\end{abstract}

\large
\end{frontmatter}
\section{Introduction}
\large
Electrostatic interaction between charged particles in aqueous electrolyte solution, together with Van der Waals interaction, plays a key role in determining physicochemical and biological properties such as the stability of colloidal suspension and interplay between biological molecules \cite%
{Derjaguin_1941, Verwey_1948, Israelachvili_2011, Russel_1989, Das_2017, Bohinc_2018}.
Electrostatic interactions in electrolyte solutions can be accurately described by electric double layer models accounting for steric effect \cite%
{Bohinc_2001, Borukhov_1997}, intra- ionic correlation \cite%
{Bohinc_2019_1, Bohinc_2019_2}, ionic hydration \cite%
{Manciu_2002, Andelman_2009, Kanduc_2014} and solvent polarization \cite%
{Bohinc_2009, Iglic_2010, Sin_2017}. 

However, those investigations have addressed mainly charged particles without any polymer coating layer \cite%
{Bohinc_2017}. Nowadays, polyelectrolyte grafted nanochannels (also called soft nanochannels) were widely studied for the development of nanofluidics achieving diode-like behavior, energy conversion and the detection of biomolecules \cite%
{Ali_2008,Ali_2009, Khatibi_2020, Khatibi_2021, Raji_2021, Das_2019}, while polyelectrolyte adsorbed particles (also called soft particles) were employed for controlling colloidal stability and drug delivery \cite%
{Mura_2013, Duval_2011_1, Duval_2015, Duval_2006, Das_2018}.  
All the studies confirmed that compared to conventional colloidal particles and nanochannels, soft colloidal particles and soft nanochannels can support excellent functions, respectively. 

A considerable amount of effort was devoted to theoretical studies of the electrostatic interaction between soft particles, accounting for various characteristics such as interpenetrating soft spheres with no particle core \cite%
{Dahnert_1994}, a salt-free medium containing only counterions \cite%
{Ohshima_2002} and the discrete-charge effect \cite%
{Richmond_1974, Richmond_1975, Nelson_1975, Miklavic_1994, Miklavic_1999, Ohshima_2014_3}, the ion size effect \cite%
{Andrews_2015}, dielectric permittivity gradient \cite%
{Duval_2011_2, Duval_2013} and the charge regulation \cite%
{Lyklema_2005},

In all the studies addressing the electrostatic interaction between two soft charged particles, the difference in Born energy between a polyelectrolyte layer and bulk ionic solution is not considered.
In fact, a long time ago, several works \cite%
{Coster_1973, Young_1998, Lopez_2003} revealed  that as the polymer layer grafted on the core has a dielectric permittivity smaller than one of the bulk ionic solution, the interactions of ions with the bound electric charges within the interface between polyelectrolyte layer and bulk ionic solution  give rise to the ion partitioning effect.

Ion partitioning effect modifies the charge density in polyelectrolyte layer and consequently affects the electrophoresis of soft charged particles as well as electrokinetics in soft nanochannels.
Very recently, ion partitioning effect  attracted a great deal of interest of researchers studying the electrophoresis of charged soft particles immersed in electrolytes \cite%
{Ganjizade_2017, Maurya_2018, Ganjizade_2018, Ganjizade_2019, Majee_2019, Gopmandal_2020, Duval_2021} and electrokinetic transport in soft nanochannels \cite%
{Suman_2016, Sadeghi_2019, Pandey_2021}. 
The analytical and numerical studies \cite%
{Ganjizade_2017, Gopmandal_2020, Duval_2021, Maurya_2018, Ganjizade_2018, Ganjizade_2019, Majee_2019}, which addressed ion partitioning effect on electrophoresis of different kinds of spherical soft particles, anticipated that ion partitioning effect can strongly modulate electrophoresis of all kinds of soft particles.  
All the studies addressing electrokinetics in a polyelectrolyte grafted nanochannel demonstrated that ion partitioning effect can improve streaming potential mediated energy conversion  \cite%
{Suman_2016} and that can significantly modulate electroosmotic transport both in soft microchannels with high grafting densities \cite%
{Sadeghi_2019} and in a nanotube containing multivalent ionic mixtures \cite%
{Pandey_2021}.

Here, it would be a natural scientific question to ask about how ion partitioning affects the electrostatic interaction between soft charged particles.
However, to the best of our knowledge, ion partitioning effect on the electrostatic interaction between two charged soft plates has never been reported in any previous studies.
Although the authors of \cite%
{Andrews_2015, Duval_2011_2} studied the disjoining pressures between soft particles or soft surfaces, their methodologies do not consider the difference in dielectric permittivity between polyelectrolyte layer and bulk ionic solution. 

In this paper, we will study impact of ion partitioning on the electrostatic interaction between two charged soft plates by using Poisson-Boltzmann approach.
We find that considering ion partitioning effect is necessary for studying the electrostatic interaction between two charged surfaces. We also illustrate that the influence of ion partitioning on electrostatic properties between charged soft surfaces depends strongly on other parameters such as the electrolyte ion size, the nature of polyelectrolyte layer, and the separation distance between the soft charged surfaces. 

\section{Theory}
\large

\begin{figure}[]
\begin{center}
\includegraphics[width=0.8\textwidth]{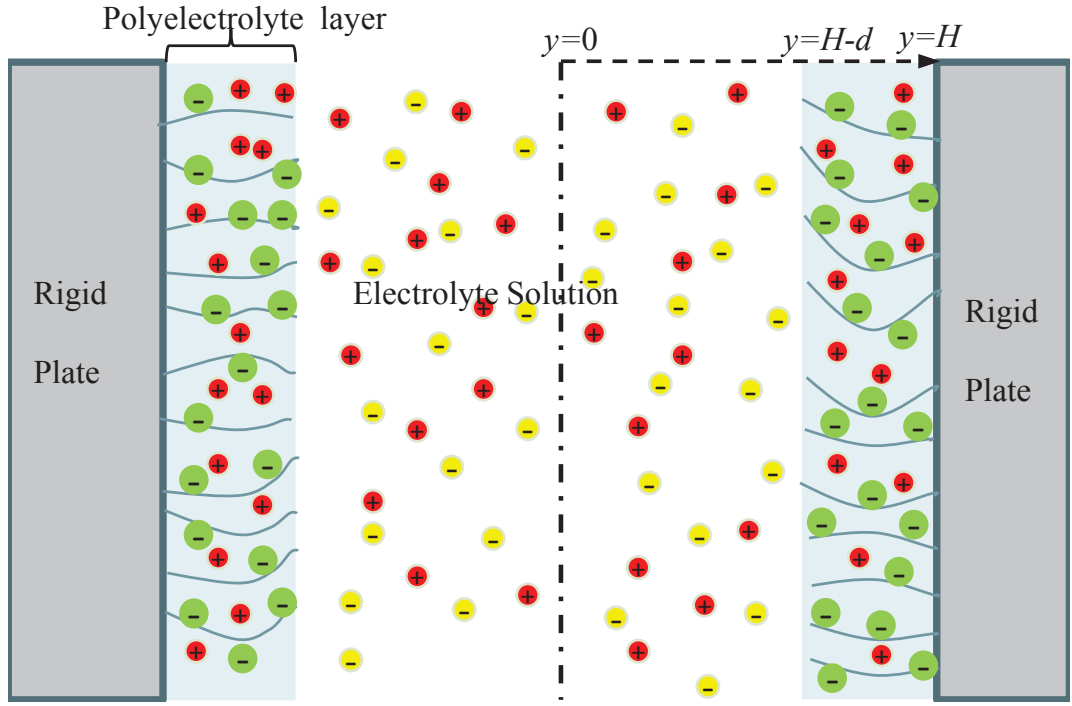}
\vspace*{8pt}
\caption{(Color online) Schematic illustration of two charged soft surfaces (consisting of a polyelectrolyte layer and a rigid plate) in contact with an electrolyte solution. The polyelectrolyte layer contains cations (shown in red) and anions (shown in yellow) as well as another kind of ions (shown in green).}
\label{fig:1}
\end{center}
\end{figure}

We consider the electrostatic interaction between two charged soft particles separated by a distance $2H$ in a binary symmetric electrolyte solution of ionic valence $z_i \left(i = +, -\right)$ and bulk ionic concentration $n_0$.  

Under the assumption that the local radius of curvature of the surface is much larger than the Debye length, we can consider the soft particles as being flat.
The geometrical symmetry of the present system allows the electric field to be set perpendicular to the particle surface. The $y$ axis is taken along perpendicular to the surface with the origin, which is located at the centerline between two charged soft particles, as shown in Fig. \ref{fig:1}. The soft particle comprises a rigid plate surrounded by an electrolyte ion penetrable polyelectrolyte layer of thickness $d$. 

The rigid plate of the particle is assumed to be neutral. The polyelectrolyte layer is considered to be charged as a uniform volume charge density due to the fixed polyelectrolyte ions with a valence $Z$ and an ionic number density $N$. The permittivity of the polyelectrolyte layer $\epsilon_p$ and the electrolyte solution $\epsilon_f$ are assumed to be different.

Since polyelectrolyte layer and bulk electrolyte solution have distinct dielectric permittivity, the self-energies of an ion, $W_i$, in the two regions differ from each other. The difference in self-energy of the ion causes ion partitioning, which can be quantified as ion partition coefficient $f_i$ with $ n_i \left(\left(H-d \right)_+ \right)/ n_i\left(\left(H-d \right)_- \right) = f_i$ \cite%
{Ganjizade_2017, Maurya_2018, Ganjizade_2018, Ganjizade_2019, Majee_2019, Gopmandal_2020, Duval_2021}
\begin{equation}
f_i  = \exp \left( { - \frac{{\Delta W_i }}{{k_B T}}} \right),
\label{eq:1}
\end{equation}
where subscript + and - refer to the right and left of region, respectively.
Here $e_0$ is the electronic charge, $k_B$ is the Boltzmann constant, $T$ is the absolute temperature, $n_i$ is the concentration of ionic species $i$ in the bulk electrolyte, respectively. The difference in Born energy of the ion, $\Delta W_i$ \cite%
{Born_1920}, is given by
\begin{equation}
\Delta W_i  = \frac{{\left( {z_i e_0 } \right)^2 }}
{{8\pi r_i }}\left( {\frac{1}
{{\varepsilon _p }} - \frac{1}
{{\varepsilon _f }}} \right),
\label{eq:2}
\end{equation}
where $r_i \left( i = 1, 2 \right)$ is the radius of ion having charge $z_i e_0$.
For simplicity, we assume that the hydrated ion radius of anions and cations is identical with each other and the radius is expressed as $r$.
Eq. (3) indicates that increasing the permittivity of polyelectrolyte layer decreases the energy difference and thus allows more electrolyte ions within the polyelectrolyte layer to exist. If the polyelectrolyte layer contains much water, the difference in permittivity between the two regions approaches to zero and finally the difference in Born energy will converge to zero. However, for the case of a typical polyelectrolyte layer with a low permittivity ratio $\epsilon_p/\epsilon_f <<1$, the Born energy difference makes electrolyte ions to hardly penetrate into the polyelectrolyte layer. 

The Poisson equation provides the relationship between the electric potential and the charge density of the system consisting of the electrolyte ions and the fixed charges within the polyelectrolyte layer. 
\begin{eqnarray}			
\nabla \left( {\varepsilon _0 \varepsilon _f \nabla \psi } \right) =  - e_0 z\left( {n_ +   - n_ -  } \right), \left| y \right| \leqslant H - d
\label{eq:3}
\end{eqnarray}
\begin{eqnarray}			
\nabla \left( {\varepsilon _0 \varepsilon _p \nabla \psi } \right) =  - e_0 z\left( {n_ +   - n_ -  } \right) - eZN,  H \geqslant \left| y \right| \geqslant H - d ,
\label{eq:4}
\end{eqnarray}
where $z=z_+=-z_-$.
When we multiply both sides of Eq. (3) by $\psi \left ( y \right)$ and rearrange the expression, we obtain the following mathematical expression for the electrostatic interaction, $P$
\begin{eqnarray}
P = \Pi  + P_{bulk}  =  - \frac{{\varepsilon _0 \varepsilon _f E\left( y \right)^2 }}
{2} + 2n_0 k_B T\cosh \left( {\frac{{e_0 z\psi \left( y \right)}}
{{k_B T}}} \right),
\label{eq:5}
\end{eqnarray}
where $\Pi$ and $P_{bulk}$  are the osmotic pressure and the electrostatic interaction for infinite distance between two soft surfaces, respectively.

As the separation between the charged surfaces reaches to the positive infinity, the electrostatic interaction is $P=P_{bulk}$.
Substituting $P_{bulk}=2 k_B T n_0$ into Eq. (5), we get the following expression,
 \begin{eqnarray}
\Pi  =  - \frac{{\varepsilon _0 \varepsilon _r E^2 \left( y \right)}}
{2} + 2n_0 k_B T\left( {\cosh \left( {\frac{{e_0 z\psi \left( y \right)}}
{{k_B T}}} \right) - 1} \right).
\label{eq:6}
\end{eqnarray}

In particular, the osmotic pressure can be expressed more simply at the centerline
\begin{equation}
\Pi  = 2n_0 k_B T\left( {\cosh \left( {\frac{{e_0 z\psi \left( {y = 0} \right)}}
{{k_B T}}} \right) - 1} \right).
\label{eq:7}
\end{equation}
			
Here, we introduce the dimensionless parameters as follows
 \begin{eqnarray}
\beta  = 1/\left( {k_B T} \right), \bar y = y/H,\bar \psi  = e_0 z\beta \psi ,\bar d = d/H,\xi  = \varepsilon _p /\varepsilon _f
\\ \nonumber K_\lambda   = \frac{{\lambda _{FCL} }}
{\lambda },\lambda  = \sqrt {\frac{{\varepsilon _0 \varepsilon _f k_B T}}
{{2n_0 e_0^2 z^2 }}}, \lambda _{FCL}  = \sqrt {\frac{{\varepsilon _0 \varepsilon _f k_B T}}
{{Ne_0^2 zZ}}}.
\label{eq:8} 
\end{eqnarray}

When we use the relations of Eq. (\ref{eq:8}), Eq. (\ref{eq:1}) and (\ref{eq:2}) can be transformed into the following dimensionless equations.
\begin{eqnarray}
\frac{{d^2 \bar \psi }}
{{d\bar y^2 }} = \frac{1}
{{\bar \lambda ^2 }}\sinh \left( {\bar \psi } \right),   1 - \bar d \geqslant \left| {\bar y} \right|
\label{eq:9} 
\end{eqnarray}
\begin{eqnarray}
\frac{{d^2 \bar \psi }}
{{d\bar y^2 }} = \frac{1}
{{\xi \bar \lambda ^2 }}\left( {\sinh \left( {\bar \psi } \right)\exp \left( { - \frac{{\Delta W}}
{{k_B T}}} \right) - \frac{1}
{{K_\lambda ^2 }}} \right),   1 \geqslant \left| \bar y \right| \geqslant 1 - \bar d,
\label{eq:10} 
\end{eqnarray}
where we assume that $\Delta W=\Delta W_+=\Delta W_-$ and  $\bar \lambda=\lambda/H$.

These equations should be solved by combining the following boundary conditions
\begin{eqnarray}
\left( {\frac{{d\bar \psi }}
{{d\bar y}}} \right)_{\bar y = 1}  = 0,\left( {\frac{{d\bar \psi }}
{{d\bar y}}} \right)_{\bar y = 0}  = 0,\\ \nonumber
\left( {\bar \psi } \right)_{\bar y = \left( {1 - \bar d} \right)_ +  }  = \left( {\bar \psi } \right)_{\bar y = \left( {1 - \bar d} \right)_ -  } ,\left( {\frac{{d\bar \psi }}
{{d\bar y}}} \right)_{\bar y = \left( {1 - \bar d} \right)_ +  }  = \left( {\frac{{d\bar \psi }}
{{d\bar y}}} \right)_{\bar y = \left( {1 - \bar d} \right)_ -  } ,
\label{eq:11} 
\end{eqnarray}
where the first condition specifies that at $y =H$, electric field strength should be zero since the rigid wall is not charged.
The second condition expresses zero electric field strength at the middle position between two homogeneous charged particles, while the third and fourth conditions reflect the continuity of the electric potential and of the normal component of the displacement vector at interface between the charged soft surface and bulk electrolyte solution.

For low potentials ($\bar \psi<<1$), Debye-H\"uckel approximation can be applied to Eq. (\ref{eq:9}) and Eq. (\ref{eq:10}), providing the following equation
\begin{eqnarray}
 \frac{{d^2 \bar \psi }}{{d\bar y^2 }} = \frac{1}{{\bar \lambda ^2 }}\bar \psi ,1 - \bar d \ge \left| {\bar y} \right| , 
\label{eq:101a}
\end{eqnarray}
\begin{eqnarray}
 \frac{{d^2 \bar \psi }}{{d\bar y^2 }} = \frac{1}{{\xi \bar \lambda ^2 }}\left( {\bar \psi \exp \left( { - \frac{{\Delta W}}{{k_B T}}} \right) - \frac{1}{{K_\lambda ^2 }}} \right),1 \ge \left| \bar y \right| \ge 1 - \bar d,
\label{eq:101b}
 \end{eqnarray}
Integrating both sides of Eq. (\ref{eq:101a}) and (\ref{eq:101b}) with consideration of boundary conditions provides the following analytical solution
\begin{eqnarray}
\bar \psi  =  B\cosh \left( {\frac{{\bar y}}{{\bar \lambda }}} \right), 1- \bar d \ge \bar y,
\label{eq:102}
\end{eqnarray}
\begin{eqnarray}
\bar \psi =\left(B_1+ C_1 \cosh \left( {\sqrt {A_1 } \bar y} \right) + D_1 \sinh \left( {\sqrt {A_1 } \bar y} \right)\right)/A_1, 1 \ge \bar y \ge 1 - \bar d,
\label{eq:103}
\end{eqnarray}
where $A_1=\frac{1}{{\xi \bar \lambda ^2 }}\exp \left( { - \frac{{\Delta W}}{{k_B T}}} \right)$ and $B_1=\frac{1}{{\xi \bar \lambda ^2 }}\frac{1}{{K_\lambda ^2 }}$,
\begin{equation}
B = \frac{{B_1 }}{{A_1 \left\{ {\cosh \left( {\frac{{1 - \bar d}}{{\bar \lambda }}} \right) - \frac{{\sinh \left( {\frac{{1 - \bar d}}{{\bar \lambda }}} \right)\left\{ {\cosh \left( {\sqrt {A_1 } \left( {1 - \bar d} \right)} \right) - \tanh \left( {\sqrt {A_1 } } \right)\sinh \left( {\sqrt {A_1 } \left( {1 - \bar d} \right)} \right)} \right\}}}{{\xi \bar \lambda \sqrt {A_1 } \sinh \left( {\sqrt {A_1 } \left( {1 - \bar d} \right)} \right)}}}\right\}}},
\label{eq:104}
\end{equation}
\begin{equation}
C_1  = \frac{{B\sqrt {A_1 } }}{{\xi \bar \lambda }}\frac{{\sinh \left( {\frac{{1 - \bar d}}{{\bar \lambda }}} \right)}}{{\sinh \left( {\sqrt {A_1 } \left( {1 - \bar d} \right)} \right) - \tanh \left( {\sqrt {A_1 } } \right)\cosh \left( {\sqrt {A_1 } \left( {1 - \bar d} \right)} \right)}},
\label{eq:105}
\end{equation}
\begin{equation}
D_1  =  - C_1 \tanh \left( {\sqrt {A_1 } } \right).
\label{eq:106}
\end{equation}
Although in \cite%
{Duval_2013}, an analytical formula for electrostatic potential was obtained within the linearized Poisson-Boltzmann approach, they considered only the surface charge density of core part, but not the difference in Born energy between soft layer and bulk ionic solution.

Considering the relations in the formula Eq. (\ref{eq:6}) for the electrostatic interaction, Eq. (\ref{eq:7}) can be expressed in dimensionless form as follows
 \begin{eqnarray}
\bar \Pi  = \frac{\Pi }
{{2n_0 k_B T}} = \left( {\cosh \left( {\bar \psi \left( {\bar y = 0} \right)} \right) - 1} \right).
\label{eq:12} 
\end{eqnarray}
Under the Debye-H\"uckel approximation, the formular is expressed as follows
 \begin{eqnarray}
\bar \Pi=0.5\bar \psi ^2 \left( {\bar y = 0} \right)
\label{eq:13} 
\end{eqnarray}

\section{Results and Discussion}

Under the boundary conditions Eq. (\ref{eq:11}), we combine Eq. (\ref{eq:9}) and Eq. (\ref{eq:10}) and solve these differential equations for $\bar \psi \left(\bar y \right)$ by using the fourth order Runge-Kutta method.

According to previous studies \cite%
{Suman_2016, Ganjizade_2017,Maurya_2018}, we have dealt with KCl aqueous solution. We have calculated the Born energy using the average hydration radii $r_{+}=r_{-}=3.3 \times 10^{-10}$m for $ K^{+}$ and $Cl^{-}$.
The bulk ionic concentration is considered to vary over the range from $10^{-1}$ mol/L to $10^{-4}$ mol/L.
The distance between soft particles is assumed to vary between H=5 and 50nm, dimensionless EDL thickness ($\bar \lambda$) can be obtained in the range of $0.1-10$.  The polyelectrolyte charge density $\rho_{fix} = ZeN$ varies in the range of $10^5$ to $10^6 C \cdot m^{-3}$ and hence the parameter $K_\lambda$ is determined as in the region of $0.1-1$.
\large
\begin{figure}[]
\begin{center}
\includegraphics[width=1\textwidth]{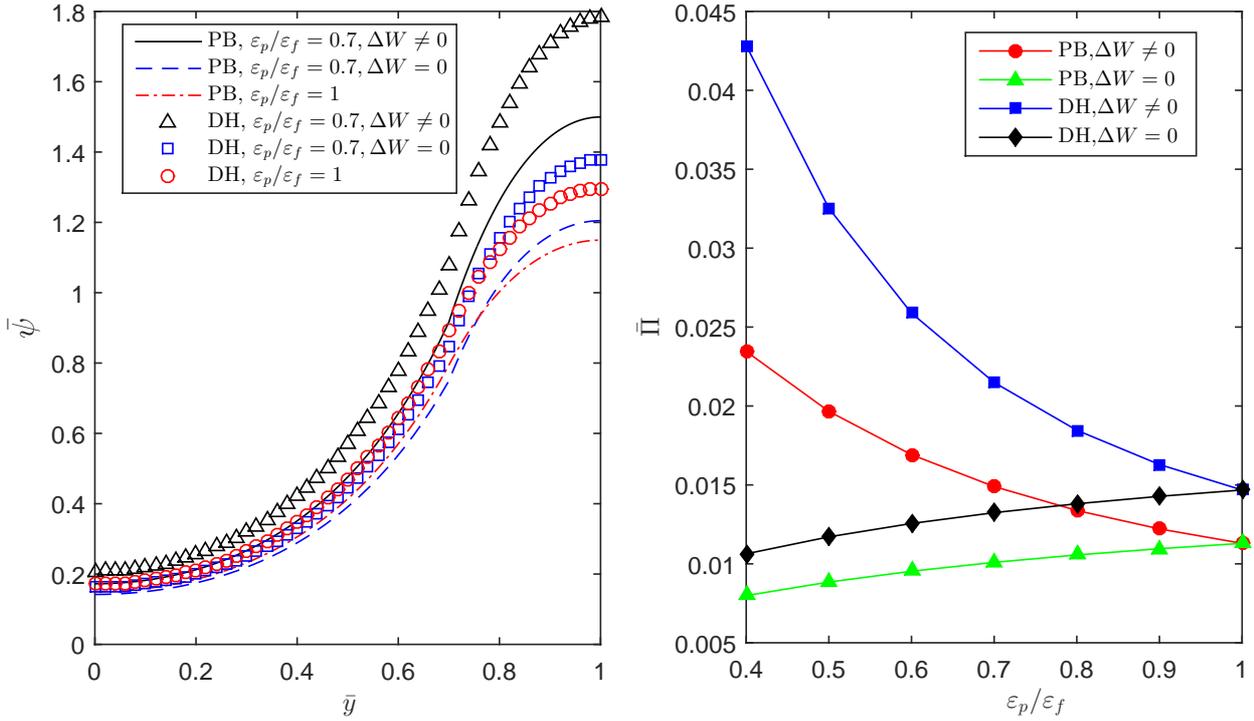}
\caption{(Color online) (a) The dimensionless electric potential as a function of dimensionless coordinate and (b) The dimensionless osmotic pressure as a function of permittivity ratio. Here $\bar d =0.3,  \bar \lambda=0.3, K_\lambda =0.7, r=3.3  \times 10^{ - 10}m.$}
\label{fig:2}
\end{center}
\end{figure}

 In calculations, we use $T = 300 K$.
We consider that the dielectric permittivity of the polyelectrolyte layer ranges from $0.4 \epsilon_f$ to $\epsilon_f$, where $\epsilon_f$  is the permittivity of the aqueous media($\varepsilon_f = 78.5$).
It is assumed that the polyelectrolyte layer thickness $d$  is $5-10nm$ and hence dimensionless thickness is $\bar d=0.3-0.6$.

Fig. \ref{fig:2}(a) shows the dimensionless electric potential profiles across the region between the two charged soft surfaces for different values of permittivity ratio by using the Poisson-Boltzmann approaches with and without Debye-H\"uckel approximation.
From the definition of ion partitioning effect, considering the effect means that we use $\Delta W \neq 0$.
The solid line, dashed line and dash-dotted line represent the results of nonlinear Poisson-Boltzmann theory with ($\Delta W \neq 0$, $\varepsilon_r/\varepsilon_f=0.7$), ($\Delta W = 0$, $\varepsilon_r/\varepsilon_f=0.7$) and ($\varepsilon_r/\varepsilon_f=1$), respectively.
The triangles, squares, circles represent the results of Debye-H\"uckel theory with ($\Delta W \neq 0$, $\varepsilon_r/\varepsilon_f=0.7$), ($\Delta W = 0$, $\varepsilon_r/\varepsilon_f=0.7$) and ($\varepsilon_r/\varepsilon_f=1$), respectively.
 
Fig. \ref{fig:2}(a) shows that for all the cases considering the ion partitioning effect ($\Delta W \neq 0$), at any position between two charged soft surfaces the electrostatic potential increases with decreasing the permittivity ratio. In particular, it is shown that the electrostatic potential within the polyelectrolyte layer is quite higher than for the case without ion partitioning effect ($\xi=1$), as already reported in Lopez-Garcia et al. \cite%
{Lopez_2003}. In fact, a lower permittivity of electrolyte solution means a lower shielding effect of it and thus yields a stronger electric force due to surface charge. On the other hand, overcoming a stronger electric field requires more energy. As a consequence, a smaller permittivity ratio $\bar \epsilon$ provides a higher electrostatic potential.   

However, when we do not consider ion partitioning effect ($\Delta W = 0$), for the cases of ($\xi < 1$),  the electrostatic potential inside the polyelectrolyte layer is higher than one for the case of ($\xi=1$), whereas the electrostatic potential outside the polyelectrolyte layer is lower rather than one for the latter case. 
The reason for the phenomenon is related to discontinuity of the electric field strength at the interface between electrolyte and polyelectrolyte layer, due to the difference in permittivity of two regions. In fact, we can easily know that at the lower permittivity of polyelectrolyte layer, the more counterions are attracted towards the polyelectrolyte layer. As a results, such a behaviour of electric potential can be understood.
\begin{figure}[]
\begin{center}
\includegraphics[width=0.6\textwidth]{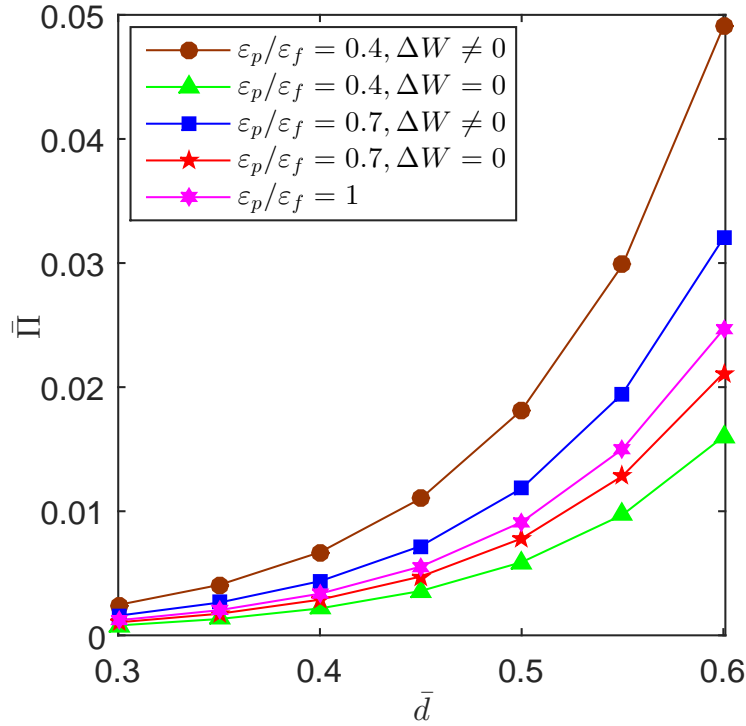}
\caption{Color online) The dimensionless osmotic pressure as a function of the dimensionless thickness of the polyelectrolyte layer. Here $\bar \lambda =0.2, K_\lambda=0.7, r=3.3 \times 10^{-10}m$.}
\label{fig:3}
\end{center}
\end{figure}
We can also observe that Debye-H\"uckel approximation allows the electric potential to be overestimated. This is due to the fact that Debye-H\"uckel approximation i.e. $\sinh\left(\bar \psi\right)\approx \bar \psi$ is valid only for cases involving low potentials. 

Fig. \ref{fig:2}(b) shows the variation of the osmotic pressure between two charged soft surfaces with the permittivity ratio. The figure clearly exhibits that for the cases of $\Delta W = 0$, the osmotic pressure increases with the permittivity ratio. The reason is that as the permittivity ratio increases ($\xi \rightarrow 1$), the electric potential at $y=0$ also increases.  
It should be also noted that for the cases of $\Delta W \neq 0$, the osmotic pressure decreases with the permittivity ratio.
This behaviour is contrary to corresponding one for $\Delta W = 0$.   Moreover,  for the cases of $\Delta W \neq 0$, osmotic pressure is higher than for corresponding cases of $\Delta W = 0$. Such a trend can be understood from the fact that Born energy increases with decreasing permittivity ratio. In fact, this can be also understood by common sense fact that the lower the permittivity of electrolyte solution, the stronger the repulsive interaction between two charged surfaces. 
 
These facts are the most important findings in this paper.  Although the authors of \cite%
{Andrews_2015} investigated a similar case, they did not consider the difference in dielectric permittivity between polyelectrolyte layer and bulk ionic solution. Consequently, they did not observe that the smaller the dielectric permittivity of the polyelectrolyte layer, the larger the difference in osmotic pressure between the cases considering or not the Born energy.
 
It is also shown that due to same reason as mentioned in Fig. \ref{fig:2}a, Debye-H\"uckel approximation allows also the electrostatic interaction to be overestimated. 

Fig. \ref{fig:3} shows numerical results for the osmotic pressure as a function of $\bar d$  for different values of the permittivity ratio for the cases.
It is shown that  increasing $\bar d$  enables the osmotic pressure to increase,  irrespective of considering or not Born energy. This fact is due to the increase of the charge density within polyelectrolyte layer. The figure also shows that like in Fig. (\ref{fig:2})b, a decrease in the permittivity ratio results in an increase in the osmotic pressure. Moreover, an increase in $\bar d$  enhances the difference in osmotic pressure between any two cases of permittivity ratio. This can be understood as follows: For a higher value of thickness of polyelectrolyte layer, the ion partitioning effect more strongly affects electric field throughout the entire region between the two soft surfaces and thus yields a higher electrostatic potential at the centerline. According to Eq. (\ref{eq:11}), a higher electric potential at the centerline provides a higher osmotic pressure.  Furthermore, we emphasize that when the thickness increases, the difference in osmotic pressure between the cases having $\Delta W=0$ and $\Delta W \neq 0$, increases.
\begin{figure}[]
\begin{center}
\includegraphics[width=0.6\textwidth]{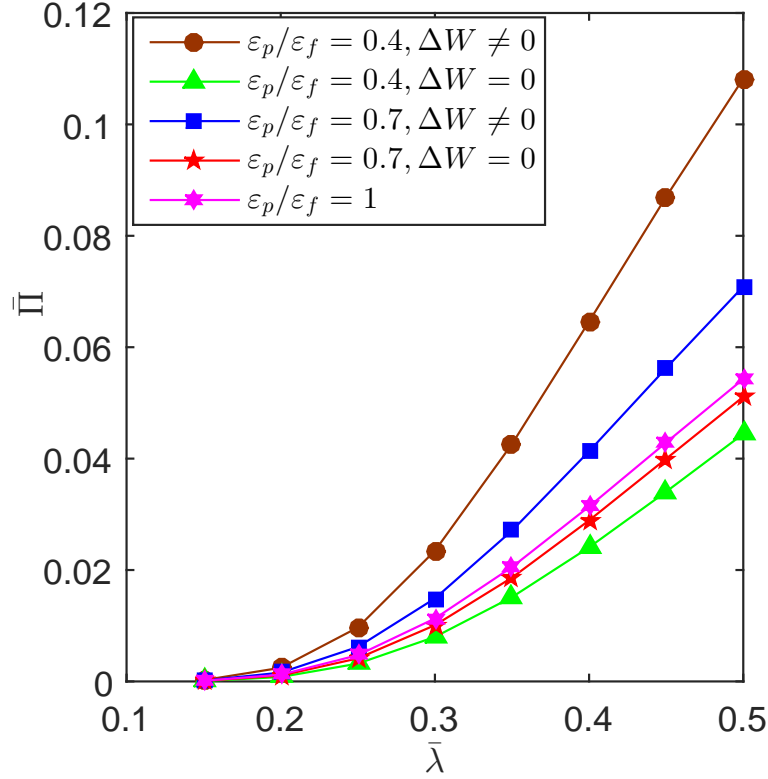}
\caption{(Color online) The dimensionless osmotic pressure as a function of the dimensionless Debye length. Here  $\bar d=0.3, K_\lambda=0.7, r=3.3 \times 10^{-10}m$.}
\label{fig:4}
\end{center}
\end{figure}
\begin{figure}[]
\begin{center}
\includegraphics[width=0.6\textwidth]{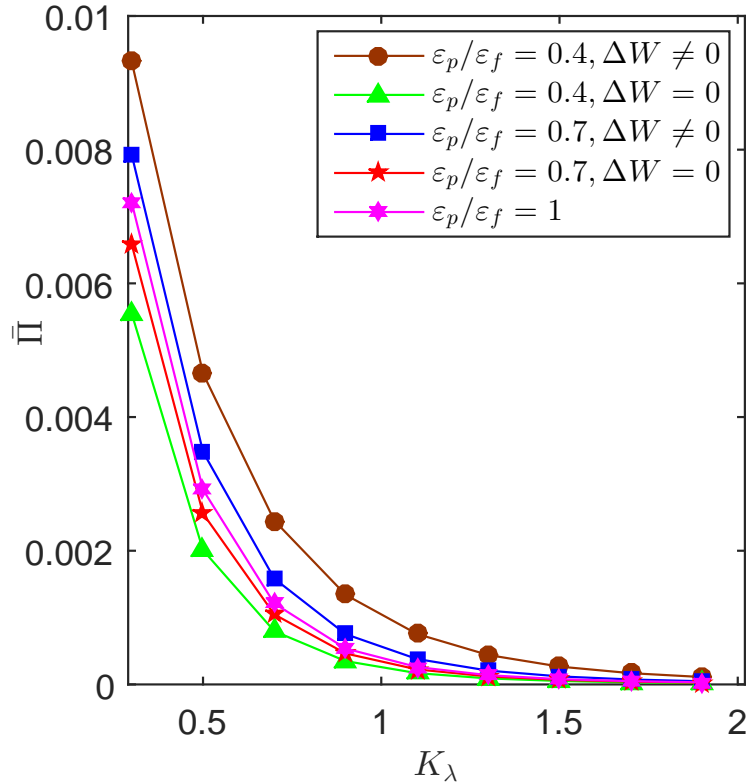}
\caption{(Color online) The dimensionless osmotic pressure $\bar \Pi$ as a function of $K_\lambda$.
Here $\bar d=0.3, \bar \lambda=0.2, r=3.3 \times 10^{-10}m$.}
\label{fig:5}
\end{center}
\end{figure}

Fig. \ref{fig:4} shows numerical results for the osmotic pressure as a function of $\bar \lambda$ at different values of the permittivity ratio for the cases with $\Delta W \neq 0$ and $\Delta W = 0$.
It is shown that increasing $\bar \lambda$ allows the osmotic pressure to increase, irrespective of considering or not Born energy. 
For the case of a higher $\bar \lambda$, ion partitioning affects a wider region from the channel wall and hence increases electrostatic potential at the centerline between two soft surfaces. As a consequence, the osmotic pressure increases with $\bar \lambda$.
Furthermore, when $\bar \lambda$ is increased, the difference in osmotic pressure for different values of permittivity ratio is increased.
In fact, it is clearly understood that when considering constant bulk ion concentration,  an increase of $\bar \lambda$ means a decrease in the separation distance between two soft surfaces increases.
Then Coulomb's law allows us to understand the above mentioned properties. In addition, it can be confirmed that an increase in $\bar \lambda$  enhances the difference in osmotic pressure between the cases having $\Delta W=0$ and $\Delta W \neq 0$.

Fig. \ref{fig:5} shows numerical results for the osmotic pressure as a function of $K_\lambda$ for different values of the permittivity ratio for the cases with $\Delta W \neq 0$ and $\Delta W = 0$.
It is shown that increasing $K_\lambda$ allows the osmotic pressure to decrease. A higher $K_\lambda$ means a lower surface charge and thus decreases the centerline potential. Consequently, increasing $K_\lambda$ results in a decrease in the osmotic pressure. 

\begin{figure}[]
\begin{center}
\includegraphics[width=0.6\textwidth]{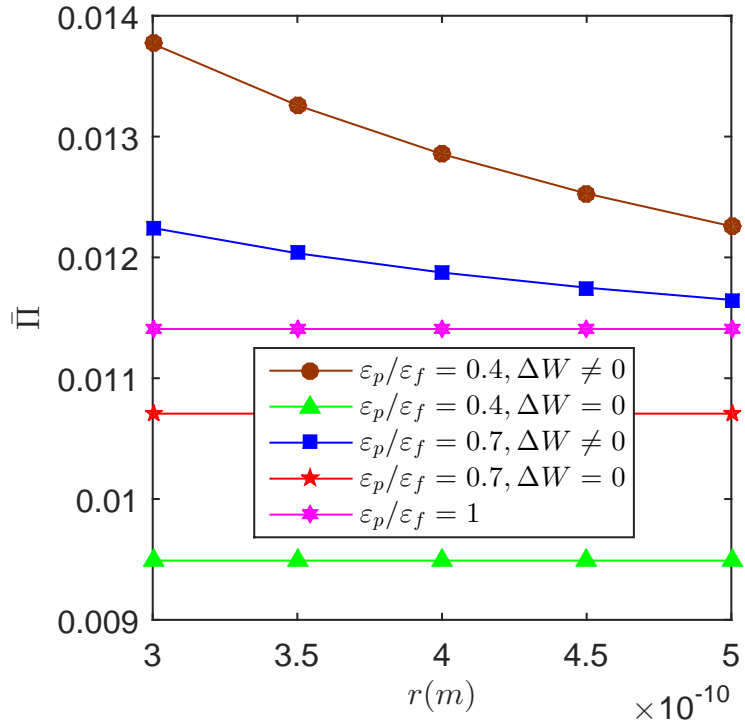}
\caption{(Color online) The dimensionless osmotic pressure $\bar \Pi$  as a function of the hydrated radius of ions in polyelectrolyte layer. Here $ \bar d=0.3, \bar \lambda =0.2, K_\lambda=0.2$.}
\label{fig:6}
\end{center}
\end{figure}

Fig. \ref{fig:6} shows the variation of the osmotic pressure as a function of hydrated radius of ions for different values of permittivity ratio.
It is shown that increasing the radius allows the osmotic pressure to decrease.
This is attributed to the fact that increasing the radius decreases Born energy and hence lowers centerline electric potential.
Furthermore, it is clearly shown that a higher ionic radius yields a smaller difference in osmotic pressure between the cases with different values of permittivity ratio. However, for the cases when not considering Born energy ($\Delta W=0$), osmotic pressures are not related with the ionic radius.

Although the present study suggests novel results, it also has some drawbacks. 
First, the present study considers only the case of hard core impermeable to both electroosmotic flow and ions. However, Maurya et al \cite%
{Maurya_2020} has shown the importance of  analytical theories for soft surface electrophoresis for soft particles with a soft permeable core.
Next, the present study does not consider surface charge density at the core material substrate and dielectric permittivity at the hard core/polymer interface. Recently, the roles of these quantities on soft surface electrostatics have been considered by Maurya et al \cite%
{Maurya_2020}.
 Moreover, the surface charge density of the core  may have profound influence on electrokinetics and electrostatic interaction \cite%
{Duval_2011_1, Duval_2015,Duval_2011_2}.
In addition, the present theory also discards the possible diffuse distribution of charged polymer segments at the grafted polyelectrolyte layer/solution interface. Theoretical and experimental investigations \cite%
{Duval_2006, Duval_2009, Duval_2010, Zimmermann_2010, Duval_2011,  Duval_2013_2, Merlin_2014, Duval_2014, Duval_2016}
accounting for diffuse distribution of charged polymer segments demonstrated that electrostatics and electrokinetics of soft particles with non-uniform charged polyelectrolyte layer are remarkably different from those of their counterparts with homogeneously distributed cationic and anionic charges. Finally, although the electric interaction between charged interfaces is affected by charge regulation of colloidal particles \cite%
{Lyklema_2005}, for the sake of simplicity, we neglect effect of charge regulation.

On the other hand, the present work should be used to evaluate measured interactions between colloidal particles, but, unfortunately, corresponding experimental data have never been reported. 
Suggesting a novel strategy to attach  nanoparticles to atomic force microscopy tips, the authors of \cite%
{Duval_2018} demonstrated that the sign of electrostatic interactions between carboxylate-terminated poly(amidoamine)
nanodendrimers and planar cysteamine-coated gold surfaces can be tuned by varying the monovalent salt  concentration in solution, under given pH conditions.  In order to verify our theory, we could try to do an experiment similar to one of \cite%
{Duval_2018}.
In the near future, we should develop a more complete theory  not only to consider the above-mentioned effects but also to well explain experimental results. 

\section{Conclusions}
	The analysis of the electric double layer mediated interaction between two charged soft plates is addressed by considering ion partitioning. The polyelectrolyte layer is considered to have different permittivity from bulk electrolyte solution. 

It is shown that when Born energy is considered ($\Delta W \neq 0$), decreasing the permittivity ratio between polyelectrolyte layer and bulk solution provides substantial increment of the osmotic pressure caused by the corresponding enhancement of the centerline potential, irrespective of the separation distance between two charged soft surfaces, the polyelectrolyte layer thickness, the surface charge density, bulk ion concentration of electrolytes. However, when Born energy is not considered ($\Delta W = 0$), the osmotic pressure is enhanced with increasing the permittivity ratio.

We find that a higher surface charge density, a longer polyelectrolyte layer thickness, a smaller separation between two surfaces and a smaller hydrated ion radius all result in an increase in osmotic pressure.
Finally, we show that a higher surface charge density, a longer polyelectrolyte layer thickness, a smaller separation between two surfaces and a smaller hydrated ion size all involve an increase of the difference in osmotic pressure between different values of permittivity ratio.

The present study could be used as a milestone for studying electrostatic interactions that are important for vast numbers of practical applications, such as drug delivery, colloidal suspension stabilization, bacterial adhesion and many more.

\end{document}